\documentclass[prd,twocolumn,floatfix,preprintnumbers]{revtex4}

\usepackage{graphicx}
\usepackage{graphicx,epsfig}
\usepackage{bm}
\usepackage{latexsym,amssymb,amsmath,float}

\newcommand {\ga} {\ {\raise-.5ex\hbox{$\buildrel>\over\sim$}}\ }
\newcommand {\la} {\ {\raise-.5ex\hbox{$\buildrel<\over\sim$}}\ }

\def\be{\begin{equation}}
\def\ee{\end{equation}}
\def\ba{\begin{eqnarray}}
\def\ea{\end{eqnarray}}

\begin{document}

\title{Reviving Quintessence with an Exponential Potential}
\author{Hui-Yiing Chang$^1$ and Robert J. Scherrer$^2$}
\affiliation{$^1$Division of Science, Mathematics and Engineering, University of South Carolina Sumter,
Sumter, SC  ~~29150}
\affiliation{$^2$Department of Physics and Astronomy, Vanderbilt University,
Nashville, TN  ~~37235}

\begin{abstract}
We examine a quintessence model with a modified exponential potential
given by $V(\phi) = V_0(1+e^{-\lambda \phi})$.  Unlike quintessence
with a standard exponential potential, our model can yield an acceptable accelerated
expansion at late times, while producing a distinct ``early dark energy" signature
at high redshift.  We determine the evolution
of the equation of state parameter, $w_\phi$, and the density parameter,
$\Omega_\phi$, as a function of the scale factor. The strongest constraints on the model come
from cosmic microwave background observations rather than supernova data.  The former give the limit
$\lambda > 13$.  This model predicts a value of the effective neutrino number
during Big Bang nucleosynthesis larger than the standard model value.
It also provides a partial solution to the coincidence
problem, in that the ratio of the quintessence energy
density is always within a few orders of magnitude of the background
radiation or matter density from the early universe up to the present,
but it does not explain why the accelerated expansion is beginning
near the present day, suggesting that these two different ways of characterizing the coincidence problem
are not entirely equivalent.
\end{abstract}

\maketitle

\section{Introduction}

Cosmological data \cite{union08,hicken,Amanullah,Union2,Hinshaw,Ade,Betoule}
suggest that approximately
70\% of the energy density in the
universe is in the form of a negative-pressure component,
called dark energy, with roughly 30\% in the form of nonrelativistic matter (including both baryons
and dark matter).
The dark energy component can be parametrized by its equation of state parameter, $w$,
defined as the ratio of the dark energy pressure to its density:
\be
\label{w}
w=p/\rho.
\ee
A cosmological constant, $\Lambda$, corresponds to the case, $w = -1$ and $\rho = constant$.

While a model with a cosmological constant and cold dark matter ($\Lambda$CDM) is consistent
with current observations,
there are many realistic models of the Universe that have a dynamical equation
of state.
For example, one can consider quintessence models, with a time-dependent scalar field, $\phi$,
having potential $V(\phi)$
\cite{Wetterich,RatraPeebles,Ferreira,CLW,CaldwellDaveSteinhardt,Liddle,SteinhardtWangZlatev}.
(See Ref. \cite{Copeland} for a review).

One of the first quintessence models to be investigated was the exponential potential,
\begin{equation}
\label{exp}
V(\phi) = V_0 e^{-\lambda \phi}.
\end{equation}
(We work in units for which $\hbar = c = 8 \pi G = 1$ throughout).
This potential arises naturally in the context of Kaluza-Klein theories, as well as in a variety
of supergravity models (see, e.g., Ref. \cite{FerreiraJoyce1998} for a discussion).
It was first explored in connection with inflation, where it produces a power-law expansion
\cite{Lucchin,Halliwell,Burd}.

Later, this potential was examined as a possible model for quintessence \cite{Wetterich,RatraPeebles,
FerreiraJoyce1998,CLW,Liddle}.  The exponential potential has the interesting property of generating
tracking solutions, i.e., for an appropriate choice of $\lambda$, the quintessence field evolves like
radiation during the radiation-dominated era, and like matter during the matter-dominated era.
This held the promise of resolving the coincidence problem, since the quintessence field
can evolve as a relatively large and constant fraction of the matter density up to the present.
However, it was soon realized that such models cannot generate the observed accelerated expansion
of the universe at late times, and they were largely abandoned.

Later, Barreiro et al. \cite
{Barreiro} attempted to resurrect the exponential quintessence
model by introducing a scalar field with a potential
given by a sum of exponentials.
Here, we investigate a simpler mechanism to allow the exponential potential to serve as a quintessence field:
an exponential potential with a nonzero offset in the potential, so that:
\begin{equation}
\label{expmod}
V(\phi) = V_0 (1+ e^{-\lambda \phi}).
\end{equation}
This model provides another example of ``early dark energy," i.e., dark energy that contributes significantly
to the expansion rate at high redshift. 

In the next section, we explore the evolution of this scalar field, and show that it gives behavior
consistent with an accelerating universe.  In Sec. III, we examine observational constraints on this model.
Our conclusions are discussed in Sec. IV.

\section{Evolution of the scalar field}

The equation of motion for
a scalar field, $\phi$, in the expanding universe
is
\be
\label{eqmotion}
\ddot{\phi} + 3H \dot{\phi} + \frac{dV}{d\phi} = 0,
\ee
where the dot indicates the derivative with respect to time, and $H$ is the Hubble parameter,
given by
\be
\label{Hubble}
H^2 = {\left ( \frac{\dot{a}}{a} \right )}^2 = \frac{\rho}{3},
\ee
where $\rho$ is the total density.  We assume a spatially flat universe throughout.
The scalar field energy density and pressure are given, respectively, by
\be
\rho_\phi={\frac{1}{2}}{\dot{\phi}}^2+V(\phi),
\ee
and
\be
p_\phi={\frac{1}{2}}{\dot{\phi}}^2 - V(\phi),
\ee
and the equation of state parameter, $w$, is given by Eq. (\ref{w}).

In the standard cosmological model (without quintessence), the density in Eq. (\ref{Hubble}) is dominated at early times
by radiation, with a density scaling as
\be
\rho_R = \rho_{R0}a^{-4}, 
\ee
while at late times it is dominated by matter, with a density given by
\be
\rho_M = \rho_{M0}a^{-3}.
\ee

In general, if the universe is dominated by a component with equation of state parameter, $w$, then the density will scale as
\be
\rho = \rho_0 a^{-3(1+w)}.
\ee
We can therefore define a ``background" equation of state parameter, $w_b$, which is given by $w_b = 1/3$ during the radiation-dominated
era, and $w_b = 0$ during the matter-dominated era.

Now consider the evolution of $\phi$ for the exponential potential given by Eq. (\ref{exp}).  For this case,
Eq. (\ref{eqmotion}) has no analytic solution.  However, it is possible to show that there is an ``attractor" toward which
the solution evolves.  For
\be
\label{limit}
\lambda^2 > 3(1+w_b),
\ee
this attractor is characterized by an equation of state
\be
w_\phi = w_b,
\ee
and a density, relative to the total density, of
\be
\label{Omegaphi}
\Omega_\phi \equiv {\frac{\rho_\phi}{\rho_\phi + \rho_b}} = {\frac{3(1+w_b)}{\lambda^2}}.
\ee
(See Refs. \cite{FerreiraJoyce1998} and \cite{CLW} for the derivation of these results).
Thus, during the radiation-dominated era, the scalar field evolves like radiation, with $w_\phi = 1/3$
and $\Omega_\phi = 4/\lambda^2$, whereas during the matter-dominated era it evolves like matter, with
$w_\phi = 0$ and $\Omega_\phi = 3/\lambda^2$.  When Eq. (\ref{limit}) is not satisfied, the attractor
is instead inflationary: the scalar field comes to dominate and $w_\phi \rightarrow -1$.

Clearly, this model cannot account for the dark energy, since observations indicate
that $w_\phi \approx -1$ at the present \cite{union08,hicken,Amanullah,Union2,Hinshaw,Ade,Betoule}.
Therefore, we modify the potential as in Eq. (\ref{expmod}).  In this model, $V_0$ must be fixed to
give the correct present-day dark energy density, leaving only a single free parameter, $\lambda$.

Using the results of Refs. \cite{FerreiraJoyce1998,CLW}, it is possible
to derive an approximate analytic prediction of the evolution of $\phi$ in this case.  At early times,
the exponential term in the potential dominates, so we have tracking behavior, with the field
evolving like radiation during the radiation-dominated era, and like matter during the matter-dominated era.
At late times, the $V_0$ term begins to become important.  To estimate the evolution in this case, we
can represent the scalar field as the sum of a constant-density part (with potential $V_0$) and
a new field, $\widetilde \phi$, which evolves in the pure exponential potential given by Eq. (\ref{exp}).
In essence, our model is identical to a quintessence field with a purely exponential potential evolving
in a $\Lambda$CDM background.

Thus, at late times, $\widetilde \phi$ tracks $w_b$ as $w_b$ evolves from $0$ to $-1$.  However, note that at the same time
Eq. (\ref{Omegaphi}) implies that $\Omega_{\widetilde \phi} \rightarrow 0$ as $w_b \rightarrow -1$.
The result is that $\rho_\phi$ scales first like matter, and then like a cosmological constant, but the evolution
is not identical to simply adding an additional dark matter component at early times and a cosmological constant
at late times.  In our model, the dark energy density decays slowly toward a constant at late times.

To see the exact evolution, we have numerically integrated Eq. (\ref{eqmotion}) with the potential
given by Eq. (\ref{expmod}) for
the sample cases $\lambda=10$, $13$ and $15$.
Since we are interested in the late-time evolution relevant for quintessence, we do not include the radiation
component.  We allow the evolution to attain the tracker solution evolution at early times, and then integrate
forward to the present day, which we define to be the scale factor at which $\Omega_\phi = 0.7$.

Fig. 1 shows the evolution of $w_\phi$ as a function of the scale factor $a$, where $a=1$ at the present.
Note that our solution is incorrect during the radiation-dominated era ($a \la 10^{-3}$), but we extend the curves
all the way back to $a=0$ for simplicity.  As expected, the equation of state parameter evolves smoothly from
$w_\phi=0$ to $w_\phi = -1$ at the present, but the details of the evolution depend on the actual value of $\lambda$.

In Fig. 2, we show the density parameter for the quintessence field, $\Omega_\phi$, as a function of $a$.  At small
$a$, the curve is nearly horizontal, as $\Omega_\phi$ is nearly constant and equal to its tracker value, while at late
times the curve evolves toward its present-day value of $\Omega_\phi \approx 0.7$.

\begin{figure}[t!]
\centerline{\epsfxsize=3.8truein\epsffile{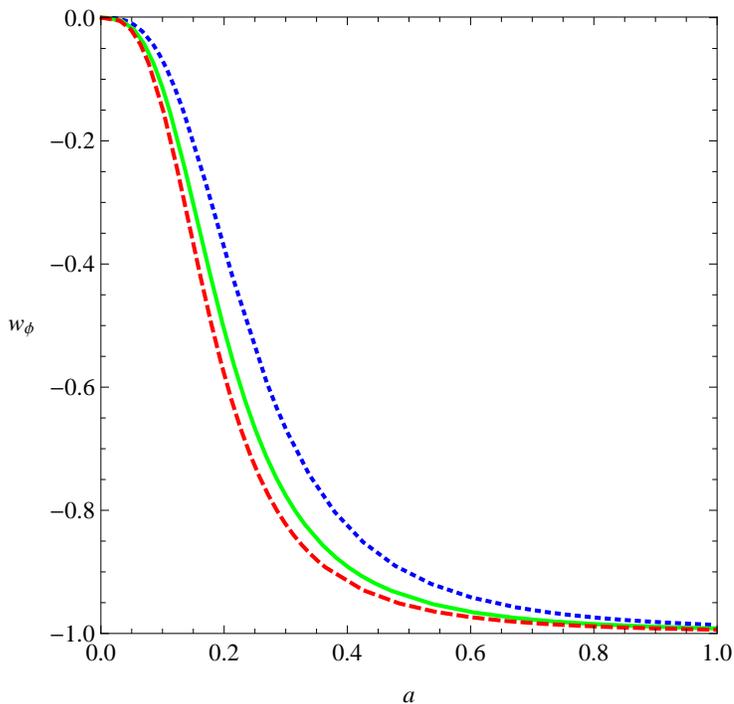}}
\caption{The evolution of the scalar field equation of state, $w_\phi$, as a function of the scale factor, $a$, where $a=1$ at the present.  
Blue dotted curve is for $\lambda=10$;
green solid curve is for $\lambda=13$;
red dashed curve is for $\lambda=15$.}
\end{figure}

\begin{figure}[t!]
\centerline{\epsfxsize=3.8truein\epsffile{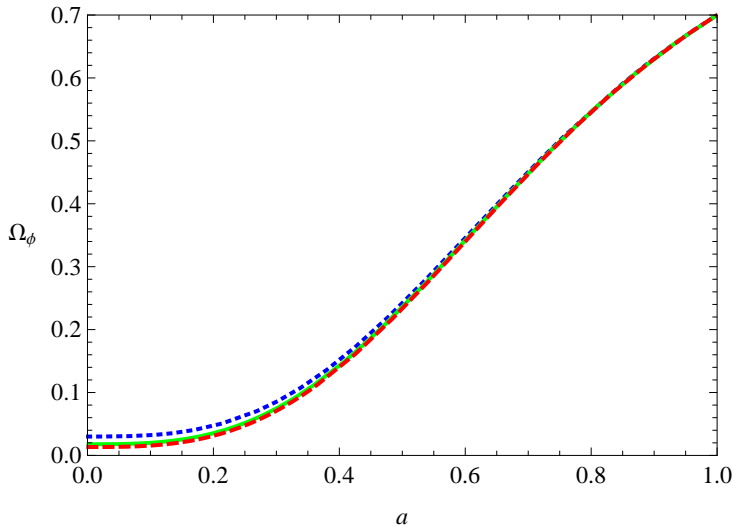}}
\caption{The evolution of the scalar field energy density parameter, $\Omega_\phi$, as a function of the scale factor, $a$, where $a=1$ at the present.  
Blue dotted curve is for $\lambda=10$;
green solid curve is for $\lambda=13$;
red dashed curve is for $\lambda=15$.}
\end{figure}

\section{Observational Constraints}

Observational data place strong constraints on models with significant early dark energy, like the one presented here.
Prior to precision CMB experiments, the best limits came from upper bounds on the energy density during big-bang
nucleosynthesis (BBN).  However, these limits have been superseded by constraints from the CMB, so the CMB constraints
are the limits we will use here.

We first note that while the density of the quintessence field evolves as radiation during the radiation-dominated era,
and as matter during the matter-dominated era, its clustering behavior is not identical to either radiation
or matter during these epochs.  The reason is that scalar fields are characterized by a sound speed of
$c_s^2 = 1$.  In contrast, cold dark matter has $c_s^2 = 0$ and radiation has $c_s^2 = 1/3$.  This
difference is most noticeable during the matter-dominated era.  Because of its large sound speed,
the scalar field does not cluster, so even a small admixture of the scalar field can produce
a distinct imprint on the CMB. 

CMB limits on additional energy density have been discussed, e.g., by Calabrese et al. \cite{Calabrese}, Samsing
et al. \cite{Samsing}, and Hojjati et al. \cite{Hojjati}.  The most useful limits for our purposes come
from Hojjati et al., who provide upper bounds on additional energy density as a function of both
redshift and sound speed, using data from Planck and WMAP9.
They parametrize the change in the expansion
rate from an additional component in terms of a parameter $\delta$, defined by
\be
H(a)^2 = \frac{\rho_{standard}(a)}{3} [1 + \delta(a)],
\ee
where $\rho_{standard}$ is the energy density in the standard $\Lambda$CDM model.

In either the matter-dominated  or radiation-dominated eras, the relation between $\Omega_\phi$ in our model
and $\delta$ in Ref. \cite{Hojjati} is given by
\be
\Omega_\phi = \frac{\delta}{1 + \delta},
\ee
so that
\be
\label{lamblimit}
\lambda = \sqrt{k \left({\frac{1}{\delta}} + 1\right)},
\ee
where $k=3(1+w_b)$.  Here, $k = 3$ during the matter-dominated era, and $k=4$ during the radiation-dominated era.

For $c_s^2 = 1$,
the constraints on $\delta$ as a function of redshift from Ref. \cite{Hojjati} are
\begin{eqnarray}
\delta &<& 0.036 ~~~ a = 10^{-4.5}, \\
\delta &<& 0.050 ~~~ a = 10^{-3.8}, \\
\delta &<& 0.160 ~~~ a = 10^{-3.4}, \\
\delta &<& 0.095 ~~~ a = 10^{-3.0}, \\
\delta &<& 0.018 ~~~ a = 10^{-1.4}. \\
\end{eqnarray}
The tightest constraints on $\delta$ occur at the lowest redshift, (largest $a$) examined in Ref. \cite{Hojjati}:
at $a = 10^{-1.4}$, $\delta < 0.018$.  This contrasts sharply with the $c_s^2 = 0$ case, for which
the additional component can simply be
absorbed into the definition of the cold dark matter density, such that the tightest constraints come
from the radiation-dominated era.  Taking $\delta < 0.018$ in Eq. (\ref{lamblimit}) gives $\lambda > 13$.
Thus, the regions in parameter space above the solid curves in Figs. 1 and 2 are ruled out.
In terms of $\Omega_\phi$, this bound translates into:
\begin{eqnarray}
\Omega_\phi &<& 0.018, ~~{\rm matter-dominated~ era},\\
\label{Ombound}
\Omega_\phi &<& 0.024, ~~{\rm radiation-dominated~ era}.
\end{eqnarray}

While the quintessence field does not behave exactly like extra radiation during the radiation-dominated era (because
it has $c_s^2 = 1$ rather than $1/3$), it is nonetheless instructive to see what energy density our limit
corresponds to in the radiation era in terms of the effective number of additional neutrinos.  In the
radiation-dominated era, the number of additional neutrinos is related to $\Omega_\phi$ as \cite{Calabrese}
\begin{equation}
\Delta N_{eff} = 7.44 \frac{\Omega_\phi}{1-\Omega_\phi}.
\end{equation}
Then our limit in Eq. (\ref{Ombound}) corresponds to
$\Delta N_{eff} < 0.18$.  Subject to this bound, our model produces a small positive value of $\Delta N_{eff}$
during Big Bang nucleosynthesis.

\section{Discussion}

It is clear from Fig. 1 that our modified exponential potential can provide a plausible model for the accelerated
expansion of the universe, with an evolution
for $w_\phi$ that differs from that of $\Lambda$CDM.
However, it predicts an evolution for $w$ that diverges only slightly from
standard $\Lambda$CDM.  For the observational bound $\lambda > 13$, the value of the equation of state parameter
at a redshift of $z=1$ ($a = 0.5$) is $w_\phi \la -0.95$, which then declines toward $w_\phi \approx -1$ at present.
In the terminology of Caldwell and Linder \cite{CL}, these are ``freezing" models.

While it will be very difficult to distinguish these models from $\Lambda$CDM using, e.g., supernova
determinations of the cosmic equation of state, these models, rather unusually for quintessence,
will actually be more strongly constrained (or confirmed) with improved CMB data.  As noted
in the previous section, a large region of parameter space is already ruled out by the Planck and WMAP9
data, so additional CMB data will either drive the allowed valued of $\lambda$ to such a large number
that the model becomes essentially indistinguishable from $\Lambda$CDM, or else show anomalies due
to additional energy density at early times.

Now consider the issue of the coincidence problem, which was one of the original motivations for introducing
quintessence with an exponential potential in the first place.  Does our modified model provide an amelioration
for the cosmic coincidence?  The coincidence problem can be stated in two different ways.  One is the fact that
the dark energy density is of the same order of magnititude as the matter density today:  $\rho_M \sim \rho_{DE}$.
This is odd because the matter density scales as $a^{-3}$, while a dark energy component derived from, e.g.,
a cosmological constant has a constant energy density.  Thus, we expect $\rho_M \gg \rho_{DE}$
at early times, and $\rho_M \ll \rho_{DE}$ in the far future, and therefore it is peculiar to find that
$\rho_M \sim \rho_{DE}$ today.  This leads to a second statement
of the coincidence problem:  the ``why now?" issue.  Why do we happen to live at a special epoch when
the dark energy density is beginning to dominate the expansion?

Our model does, to some extent, ameliorate the coincidence problem when it is expressed in terms of energy
densities.  At early times, the quintessence density tracks the radiation and matter densities,
such that from the early universe up to the present, the quintessence energy density can be a nonnegligible fraction
($\sim 0.02$) of the background
radiation or matter density, rather than being many orders of magnitude smaller.
On the other
hand, the model does nothing to answer the question of ``why now?"  The model parameters must still be tuned
so that the $V_0$ term in Eq. (\ref{expmod}) begins to dominate the $V_0 e^{-\lambda \phi}$ term at around
the present day.  Perhaps the most interesting result is that this model shows that these two ways of expressing
the coincidence problem may not be, as is usually assumed, entirely equivalent.  It is possible to construct a
model (i.e., this one) for which the densities of the dark energy and matter are not widely separated over much of the early
universe, but which still retains the need for us to live in a ``special" epoch, when the acceleration is just beginning.
(For a related, but different approach to this question, see Ref.
\cite{Fedrow}).

\section{Acknowledgments}
R.J.S. was supported in part by the Department of Energy
(DE-FG05-85ER40226).

\end{document}